\documentclass[twocolumn,secnumarabic,amssymb, nobibnotes, aps, pra]{revtex4-2}
\usepackage{color}

\newcommand{\vect}[1]{{\mbox{\boldmath $#1$}}}

\usepackage{amsmath,amssymb}
\usepackage{bm}
\usepackage{graphicx}

\begin{document}

\title{A fiber-type optomechanical array using high-Q microbottle resonators}

\author{Motoki Asano}
\author{Hiroshi Yamaguchi}
\author{Hajime Okamoto}

\affiliation{NTT Basic Research Laboratories, NTT Corporation, 3-1 Morinosato Wakamiya, Atsugi-shi, Kanagawa 243-0198, Japan.}
\date{\today}

\begin{abstract}
We demonstrate a fiber-type optomechanical array consisting of elastically interconnected silica microbottle resonators with high-Q optical and mechanical modes. In total, fifty optomechanical resonators fabricated by fine glass processing are uniformly arrayed on a silica fiber. Evanescent coupling of a tapered optical fiber to an arbitrary resonator  allows for highly sensitive readout and efficient actuation of mechanical motion at an arbitrary position in the array. Phonon propagation through the fifty microbottles is achieved by both linearly and parametrically driving a mechanical mode at one end and by detecting it at the other end. This optomechanical array is scalable, tunable, and lithography-free and can be extended to fiber-based sensory applications with structural flexibility and operability in various environments.
\end{abstract} 
\maketitle

\section{Introduction}
Cavity optomechanics offers highly sensitive readout and efficient driving of mechanical motion via an optomechanical coupling between cavity photons and phonons \cite{aspelmeyer2014cavity}. Recently, multiple cavity optomechanical devices that have optical and or elastic couplings have been developed \cite{massel2012multimode,lee2015multimode,ruesink2016nonreciprocity,nielsen2017multimode}. Such a cavity optomechanical array has three distinct advantages: (i) scalable expansion of the number of photonic and phononic modes; (ii) interconnection between the spatially distant resonator nodes with energy transfer; and (iii) appearance and control of non-trivial many-body effects (e.g., non-reciprocal transmission). The arrays are applicable to multimode optomechanical sensors, information routing of photons and phonons, and fundamental studies based on analogies to many-body systems, such as PT symmetry \cite{jing2014pt,xu2016topological,zhang2018phonon}, Anderson localization \cite{garcia2017optomechanical,arregui2019anderson,arregui2023cavity}, artificial gauge \cite{fang2017generalized,mathew2020synthetic}, and topological propagation \cite{ren2022topological,slim2023optomechanical}.

Cavity optomechanical arrays have so far been demonstrated using an on-chip optomechanical resonators, such as photonic-phononic crystal cavities \cite{ren2022topological,slim2023optomechanical} and whispering gallery mode (WGM) optomechanical resonators \cite{zhang2015synchronization}. Sophisticated semiconductor fabrication techniques have been used to make photonic-phononic crystal cavities through optical and elastic couplings. Arrays of on-chip WGM optomechanical resonators, such as microdisks, have been built by utilizing strong photon-photon couplings via the optical evanescent field. Various demonstrations have been reported, such as on phononic topological propagation \cite{ren2022topological,slim2023optomechanical} and mechanical synchronization \cite{zhang2015synchronization}, but the fixed-by-design on-chip architecture cannot be extended to on-demand applications, such as externally tunable devices \cite{pollinger2009ultrahigh,zhou2017broadband} and free-access sensors \cite{asano2022free,asano2023cavity}.

In this article, we focus on another candidate, i.e. fiber-type optomechanical resonators, such as microspheres and microbottles. The fiber-type systems have the following outstanding properties that on-chip devices do not have: lithography-free formability, structure flexibility, and operability in various environments, while maintaining their excellent intrinsic optomechanical performance \cite{macdonald2016optomechanics,asano2016observation}. Coupled array structures are expected to enable novel functions in fiber-based technology.

Here, we built a fiber cavity optomechanical array, referred to as a chained-microbottle resonator (CMBR), consisting of elastically coupled microbottle resonators. The CMBR includes fifty microbottle resonators in total and is fabricated through fine glass processing. Each microbottle has an optomechanical coupling between high-Q optical whispering gallery modes (WGMs) and high-Q mechanical radial-breathing modes (RBMs) via radiation pressure and can be used for highly sensitive readout and efficient actuation of mechanical motion [see Fig. 1(a)]. The optomechanical coupling in the CMBR is selectively activated at an arbitrary node by injecting laser light into individual WGMs by utilizing a tapered optical fiber. While the optical WGMs are locally confined in each microbottle, the mechanical RBMs are coupled with the adjacent nodes because of their spatial modes overlap along the fiber axial directions. Thus, owing to the mechanical couplings, optically-driven mechanical motion at a resonator is transferable along the one-dimensional array and can be sensitively detected at the arbitrary resonator through the optomechanical coupling. When the mechanical couplings among all of the nodes are in the strong coupling regime, phonon propagation between the two resonators at the ends becomes possible [see Fig. 1(b)]
\begin{figure}[htbh]
\includegraphics[width=8.5cm]{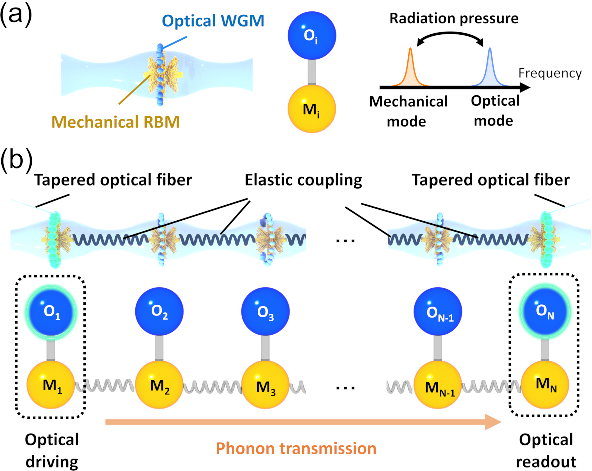}
\caption{\label{fig1} (a) Schematic diagram of a microbottle resonator, which is an element of the CMBR, with an optomechanical coupling between the optical WGM and mechanical RBM. (b) Schematic diagram of the CMBR including optomechanical and elastic coupling. Laser light (colored green) is guided into an arbitrary microbottle by using a tapered optical fiber. By combining optical driving and readout of the mechanical mode ({\it e.g.,} at the 1st and $N$th nodes, respectively), phonon propagation over the CMBR is possible. }
\end{figure}

\section{Experiment}
\subsection{Device}
The CMBR was fabricated on a silica optical fiber and included fifty microbottle nodes ($N=50$) made by periodically forming fifty-one bottlenecks via the heat-and-pull technique \cite{spillane2003ideality}. To maintain the uniformity of the microbottle structures, the tension between the ends of the silica fiber was monitored before pulling to form each neck. Figure 2(a) shows an optical microscope image of the whole structure and Fig. 2(b) shows an enlargement including five microbottle structures. Each node consists of a 125-$\mathrm{\mu m}$-diameter bottle with about $115$-$\mathrm{\mu m}$-diameter necks and neck-to-neck length of about $700$ $\mathrm{\mu m}$ (i.e., the whole device length is about 3.5 cm). 

The optical resonance of the CMBR was observed by using the probe setup shown in Fig 2 (c). An external cavity dipole laser (ECDL) operating at wavelengths around 1550 nm was used for probing the optical transmission spectrum by scanning its frequency with appropriate optical power ($\sim$1 mW) and polarization. All measurements were made under atmospheric conditions. The optical transmission spectra in six microbottles (1st, 2nd, 3rd, 48th, 49th, and 50th) were measured by adjusting the contact position of the tapered optical fiber with the CMBR. Figure 2 (d) shows typical optical Q factors in each microbottle. The average optical Q factors among the six resonators reached $\bar{Q}_\mathrm{opt}=8.3\times 10^6$, which is comparable to a standard WGM resonator. 

These high-Q optical resonances provide high displacement sensitivity, and they allowed us to probe the thermal fluctuations of the mechanical RBMs. Note that the probe laser frequency was thermally locked in the middle of the slope of cavity resonances \cite{carmon2004dynamical}. The power spectral density at the 50th node is shown in Fig. 2(e). The power spectra of the thermal fluctuations can be utilized for estimating the mechanical Q factors and the vacuum optomechanical coupling constants, defined as $g_{\mathrm{OM},j}\equiv x_{\mathrm{zpf},j}\partial \omega_{\mathrm{opt},j}(x) /\partial x $ where $\omega_{\mathrm{opt},j}$ is the optical frequency as a function of the mechanical displacement $x$, and $x_{\mathrm{zpf},j}$ is the zero-point fluctuation in the mechanical mode ($j=1,2,3,48,49,50$). Here, they were evaluated in the lowest frequency mode as a representative example. The mechanical Q factors were over $10^3$, and their average value reached $\bar{Q}_\mathrm{mech}=4.8\times 10^3$, which is comparable to standard WGM optomechanical resonators [see Fig. 2(f)]. To calibrate the optomechanical phase modulation, a calibrated phase modulation tone generated by an optical phase modulator (PM) was measured simultaneously. The average vacuum optomechanical coupling constant among the six resonators reached $\bar{g}_{\mathrm{OM}}/2\pi=246 $ Hz [see Fig. 2(g)]. The single optomechanical node thus acted like a high-Q optomechanical resonator in the same way as the previously reported single- \cite{macdonald2016optomechanics,asano2016observation} and twin-microbottle structures \cite{asano2022free,asano2023cavity}.

Multiple thermal fluctuation peaks appeared in the PSD, and they originated from two different sources. One source is the multiple axial modes in the microbottle structure, whose frequency separations are several hundred kHz. The other is the elastic couplings among the adjacent nodes, whose frequency separations are several tens of kHz. The frequency separation corresponds to the elastic coupling strength, which is larger than the mechanical linewidth. For instance, the double peaks around 31 MHz in Fig. 2(e) reflect the elastic coupling strength $g_\mathrm{M}/2\pi=42.8$ kHz, which is about ten times larger than the mechanical dissipation.

\begin{figure}[htb]
\includegraphics[width=8.5cm]{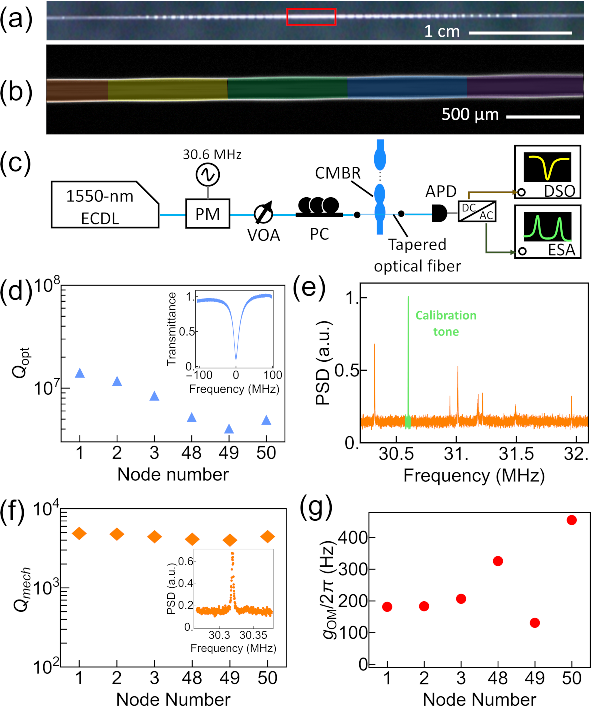}
\caption{\label{fig2} (a) Optical microscope image of the whole structure of the CMBR. (b) Optical microscope image with a focused view of the CMBR around the red squared area in (a). Each microbottle is shaded in different colors. (c) Schematic diagram of the experimental setup for evaluating optical properties and observing the thermal fluctuation of the mechanical modes in each microbottle. The abbreviations are as follows: variable optical attenuator, VOA; polarization controller, PC; avalanche photodiode, APD; optical phase modulator, PM; digital sampling oscilloscope, DSO; electric spectrum analyzer, ESA.(d) Optical Q factors obtained in the 1st, 2nd, 3rd, 48th, 49th, and 50th microbottles. The inset shows a typical transmission spectrum obtained in the 1st microbottle. (e) Power spectral density (PSD) of the thermal fluctuation of the mechanical modes and the calibration tone (at 30.6 MHz) in the 1st microbottle. The calibration tone for estimating $g_\mathrm{OM}$ is colored green. (f) Mechanical Q factors obtained in the six microbottles. (g) Vacuum optomechanical coupling constants obtained in the six microbottles.}
\end{figure}

\subsection{Phonon transmission spectrum}
Owing to existence of elastic couplings between the adjacent nodes, the mechanical vibration driven via optomechanical coupling is guided over the CMBR between the ends. Here, we contacted the tapered optical fiber on the 1st and 50th nodes for optical driving and detection of mechanical vibration, respectively. The phonon transmission spectrum was obtained by driving the mechanical vibration at the 1st node by using intensity-modulated light from an intensity modulator (IM) and by detecting the propagated vibrations at the 50th node by using a probe laser [see Fig. 3(a)]. The power of the drive and probe laser were set to about 10 mW and 1 mW, respectively. Figure 3(b) shows the phonon transmission spectra ($S_{21}$ signal) monitored with a vector network analyzer (VNA) in which the RF drive port was connected to the IM for the drive laser and the detection port was connected to the photodiode for the probe laser. 

Figure 3(b) shows the phonon transmission spectra in the driven measurement. Whereas the thermal fluctuation measurement reflected the local projection of the coupled mechanical modes, this driven measurement unveils the transfer function in the CMBR. Thus, the multiple peaks in the spectra are evidence of the elastic interconnection between the two ends. To clarify the contribution from the higher-order axial modes, we performed a finite-element-method (FEM) simulation by using COMSOL Multiphysics (see Appendix A). The insets in Fig. 3(b) show the displacement profile of each axial mode calculated with periodic boundary conditions. Because the higher-order axial modes have large mode overlaps between the adjacent nodes, the mechanical couplings in those modes (31-32 MHz) become stronger than those in the lower-order axial modes (around 30.2 MHz). This caused the transmitted amplitude in the higher-order axial modes to be larger than those of the lower-order ones. The CMBR thus operated as a phonon waveguide in which the vibrating energy is transferred between the two ends. 

\begin{figure}[htb]
\includegraphics[width=8.5cm]{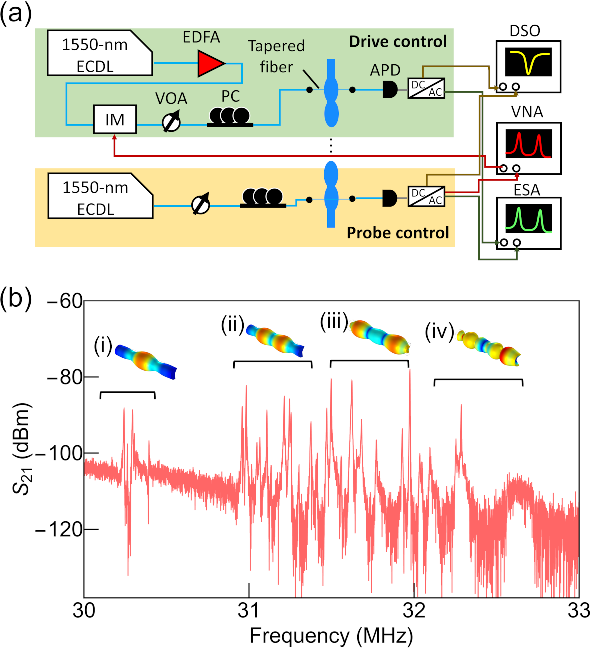}
\caption{\label{fig3} (a) Schematic diagram of the experimental setup for phonon propagation in the CMBR. The abbreviations are the same as that in Fig. 2 with additional ones as follows: erbium-doped fiber amplifier, EDFA; intensity modulator, IM; vector network analyzer, VNA. (b) A phonon transmission spectrum in the CMBR while optomechanical driving and detection are performed at both ends. The insets show the spatial profile of the radial breathing modes with the (i) 1st-, (ii) 2nd-, (iii) 3rd-, (iv) 4th- order axial mode numbers calculated in the finite-element-method simulation (see Appendix A). }
\end{figure}

\subsection{Parametric oscillation and its propagation}
High-Q optomechanical resonators provide optomechanical parametric oscillations whereby the extremely narrow linewidths in the self-oscillation phonon states can be used in sensor applications \cite{liu2013sub,yu2016cavity,guha2020force}. Since the oscillation threshold is proportional to $Q_\mathrm{opt}^{-2}Q^{-1}_\mathrm{mech}$ \cite{kippenberg2005analysis}, the high-Q CMBR can efficiently induce parametric oscillation and transmit self-oscillating phonon states through elastic coupling. The optomechanical coupling in each microbottle provides an optomechanical gain, $\Gamma_\mathrm{OM}$, by using a blue-detuned laser whose detuning is close to the sum of the optical and mechanical resonance frequencies. When the optomechanical gain is larger than the total dissipation, a mechanical mode starts to oscillate. Because the optomechanical coupling is localized in each resonator, the parametric optomechanical oscillation occurs in a mechanical mode in the single microbottle in contact with the tapered optical fiber. This local parametric oscillation becomes more dominant as the number of nodes increases (see Appendix B). 

Owing to such a local parametric oscillation, the oscillating phonon at the 1st node propagates to the 50th node via mechanical coupling [Fig. 4(a)]. Here, the threshold of the local parametric oscillation was evaluated by setting the laser frequency in the blue-detuned regime for both oscillating and probing the mechanical mode at the 1st node. The threshold optical power of $12.2$ mW was determined from the input optical power dependence [Fig. 4(b)]. The oscillation amplitude in Fig. 4(b) was estimated through a fast Fourier transformation while scanning the laser frequency. Propagation of the mechanical oscillation was demonstrated by using two tapered optical fibers for pumping and probing at the 1st and the 50th node, respectively. Figures 4(c) and (d) show the PSD observed in the 1st node and the 50th node, respectively, with an optical pump power of $P_\mathrm{D}=20.3$ mW [the optical power at the probe node is ten times smaller than the drive (about 200 $\mathrm{\mu W}$)]. The mechanical oscillation signal was observed in the 1st and the 50th nodes, which is evidence of phonon propagation of the local parametric oscillation signal. In addition to the signal in the frequency domain, the oscillating periodic signals in the time domain at the 1st and 50th nodes reflected the conservation of signal coherence through the phonon propagation [see Fig. 4(e) and (f)]. 

\begin{figure}[htb]
\includegraphics[width=8.5cm]{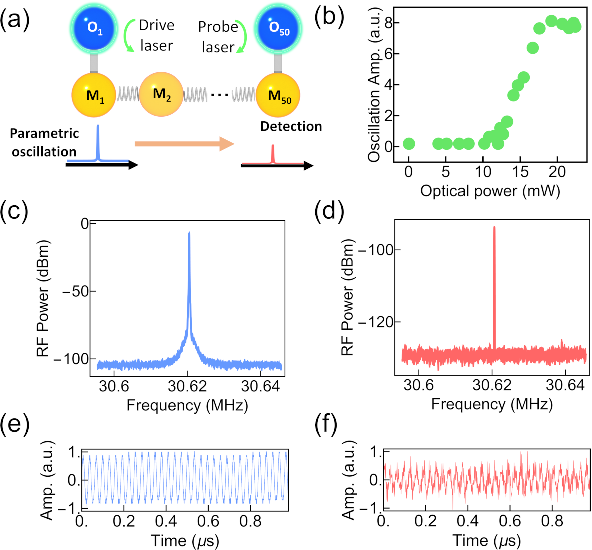}
\caption{\label{fig4}(a) Schematic diagram of phonon propagation with parametric optomechanical oscillation (b) Mechanical oscillation amplitude with respect to the input optical power at the 1st node. The parametric oscillation occurs at a threshold optical power of $12.2$ mW. (c, d) RF spectra of optomechanical parametric oscillation observed at the 1st node (c) and 50th node (d). (e, f) Time domain signals of optomechanical parametric oscillation observed at the 1st node (e) and 50th node (f).}
\end{figure}

\section{Discussion}
The CMBR was fabricated by forming a microbottle structure one-by-one, which is a completely different fabrication process from that of on-chip coupled mechanical resonators (e.g., wet etching process). The geometry of each element is tunable by sequentially fabricating and evaluating each microbottle \cite{sumetsky2012snap}. This adaptive fabrication process will not only decrease the deviations in the mechanical properties of microbottles such as the resonance frequency and linewidth, but also introduce more functional phononic structures, such as point defects \cite{hatanaka2014phonon} and zero-dimensional topological edges \cite{xiao2015geometric,kim2017topologically,esmann2018topological}.

The simple and selective laser input by using tapered optical fibers allows us to increase the number of control cavity nodes. Although this proof-of-principle experiment was done by simultaneously injecting the laser light into two microbottle resonators, utilizing more laser inputs by using multiple tapered optical fibers can be used to perform more sophisticated phonon routing by multiply combining optomechanical cooling, amplification, and nonlinearity \cite{monifi2016optomechanically,navarro2017nonlinear,leijssen2017nonlinear,zhang2021optomechanical}.

Furthermore, the fiber-type geometry of the cavity optomechanical array enables us to readily use the device in various environments, e.g., in liquids, similarly to our previously reported twin-microbottle resonator \cite{asano2022free,asano2023cavity}. This would offer new sensory applications in opto-mechano-fluidics \cite{bahl2013brillouin,gil2015high,gil2020optomechanical,asano2022free,asano2023cavity} by utilizing multiple mechanical modes for spatial resolution with highly sensitive optical readout.

\section{Conclusion}
We demonstrated a fiber-type optomechanical array using a chained-microbottle resonator. The array consisted of fifty high-Q microbottle resonators that were elastically interconnected through fine glass processing. Phonon propagation over the CMBR was demonstrated by combining optical linear (and parametric) driving at one end of the array and sensitive detection of mechanical motion at the other end. The high scalability, accessibility of each resonator, and controllability with a high-Q optical cavity will expand the applications of optomechanical arrays with unique fiber properties, such as structural flexibility and operability in various environments.

\section*{Acknowledgements}
This work was supported by JSPS KAKENHI (21H01023, 23H05463).

\appendix 
\section{Finite-element-method simulations}
To examine the mechanical axial mode distributions, we calculated the mechanical eigen-modes in our CMBR structure by utilizing a finite-element-method (FEM) simulator (COMSOL Multiphysics). Here, periodic boundary conditions were imposed on a microbottle structure (maximum diameter of 125 $\mathrm{\mu m}$, neck diameter of $115$ $\mathrm{\mu m}$, and neck separation length of $700$ $\mathrm{\mu m}$) so as to achieve the mechanical band structure. As shown in Fig. S1, different modes with respect to the axial mode number appear in each mechanical band. In contrast to the higher-order axial modes, the lower-order axial modes are strongly confined around the center of the microbottle structure. This implies that the lower-order axial modes have a smaller elastic coupling than the higher ones do because of the small internode mode overlap, which causes the local parametric oscillation discussed in the main text. 

\begin{figure}[htb]
\includegraphics[width=8cm]{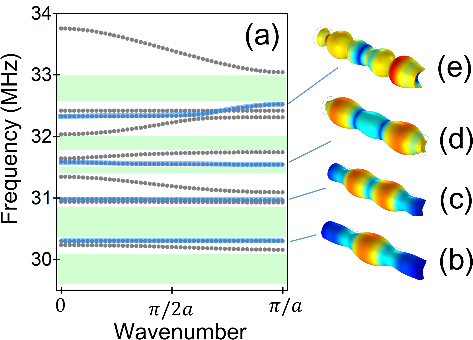}
\caption{\label{figS2}{\bf Mechanical band structure calculated by the FEM simulation.} (a) The mechanical band structure for different axial-order RBMs colored blue. The green shaded areas correspond to the band gap. (b)-(e) Spatial distributions of the 1st- (b), 2nd- (c), 3rd- (d), and 4th-order (e) of the axial number. }
\end{figure}

\section{Local parametric oscillation}
Here, we consider local parametric oscillation at the 1st node (i.e., an edge node in the CMBR) where the other coupled $N-1$ mechanical nodes act as a dissipation channel. To simplify the discussion, we assume a uniform mechanical frequency $\Omega_\mathrm{M}$, damping rate $\Gamma_\mathrm{M}$, and coupling strength $g_\mathrm{M}$. Owing to the optomechanical coupling at the 1st node, the oscillation condition can be expressed as
\begin{align}
\Gamma_\mathrm{OM}>\Gamma_\mathrm{M}+\Gamma_\mathrm{C},
\end{align}
where $\Gamma_\mathrm{OM}$ is the optomechancial gain and $\Gamma_C$ is the rate of dissipation to the other coupled nodes. Here, we have taken into account the following coupled mode equations:
\begin{align}
&\ddot{x}_1+\Gamma_\mathrm{M} \dot{x}_1+\Omega_\mathrm{M}^2 x_1+g_\mathrm{M} \Omega x_2 = F_\mathrm{OM},\\
&\ddot{x}_2+\Gamma_\mathrm{M} \dot{x}_2+\Omega_\mathrm{M}^2 x_2+g_\mathrm{M} \Omega x_1+g_\mathrm{M}\Omega x_3 = 0,\\
&\vdots \nonumber \\
&\ddot{x}_N+\Gamma_\mathrm{M} \dot{x}_N+\Omega_\mathrm{M}^2 x_N+g_\mathrm{M} \Omega x_{N-1} = 0,
\end{align}
where $x_i$ is the mechanical displacement of node $i$, and $F_\mathrm{OM}$ is the optomechanical force exerted on the 1st node. Note that we ignore the Langevin force inducing thermal fluctuations. The mechanical displacements $(x_2, x_3, \cdots x_N)$ are regarded as a dissipation channel (i.e., waveguide mode) by transforming to the waveguide basis $(w_2,w_3\cdots w_{N})$ given by
\begin{align}
w_k=\sqrt{\frac{2}{N}}\sum_{l=2}^{N} \sin\left(\frac{(k-1)(l-1)\pi }{N}\right) x_l,
\end{align}
where the factor of $\sqrt{2/(N+1)}$ is derived from the requirement for a unitary transformation (i.e., energy conservation). Importantly, the inverse transformation is equivalent; i.e., replacement of $w_i$ by $x_i$ gives a similar form. Thus, by taking the Fourier transform, we can rewrite the coupled mode equations in the frequency domain,
\begin{align}
&\chi_\mathrm{M}(\omega)x_1(\omega)+\frac{\sqrt{2}g_\mathrm{M}\Omega}{\sqrt{N}} \sum_{l=2}^{N}\sin\left(\frac{l\pi}{N}\right)w_l(\omega)= F_\mathrm{OM},\label{eq1}\\
&\left[\chi_\mathrm{M}(\omega)+2g_\mathrm{M}\Omega \cos\left(\frac{(k-1)\pi}{N}\right)\right] w_k(\omega)\nonumber\\
&\hspace{20pt}+\frac{\sqrt{2}g_\mathrm{M}\Omega}{\sqrt{N}}\sin\left(\frac{(k-1)\pi}{N}\right)x_1(\omega)= 0,\label{eq2}
\end{align}
where $\chi_\mathrm{M}(\omega)\equiv (\Omega^2-\omega^2)-i\omega\Gamma$ and the factor $2g_\mathrm{M}\Omega \cos\left(\frac{(k-1)\pi}{N}\right)$ is derived as the eigenvalue of the corresponding waveguide mode. By substituting Eq. (\ref{eq2}) into Eq. (\ref{eq1}), the dynamics of $x_1$ become
\begin{widetext}
\begin{align}
\left[\chi_\mathrm{M}(\omega)+\frac{2}{N+1}\sum_{l=1}^{N-1}\frac{g_\mathrm{M}^2\Omega^2}{\left(\chi(\omega)+2g_\mathrm{M}\Omega\cos\left(\frac{l\pi}{N}\right)\right)}\sin\left(\frac{(l+1)\pi}{N}\right)\sin\left(\frac{l\pi}{N}\right)-F_\mathrm{OM}\right]x_1(\omega)=0.
\end{align}
\end{widetext}
Thus, the oscillation condition is achieved by taking account of the imaginary part of the total susceptibility, i.e.,
\begin{align}
\Gamma_\mathrm{M}+\sum_{l=1}^{N-1}\Gamma_{\mathrm{C}(l+1)}<\Gamma_\mathrm{OM} \label{eq3},
\end{align}
where we have defined $\mathrm{Im}[F_\mathrm{OM}]=\omega\Gamma_\mathrm{OM}$ and $\Omega_l\equiv 2g_\mathrm{M}\cos\left(\frac{l\pi}{N}\right)$. The left-hand sides in Eq. (\ref{eq3}) can be further reduced to
\begin{widetext}
\begin{align}
&\Gamma_\mathrm{M}+\frac{g_\mathrm{M}^2\Omega^2}{N+1}\sum_{l=1}^{N-1}\frac{\Gamma_\mathrm{M}}{(\omega^2-\Omega^2+\Omega\Omega_l)^2+\omega^2\Gamma_\mathrm{M}^2}\left[1-\cos\left(\frac{(2k+1)\pi}{N}\right)\right]\\
&\approx \Gamma_\mathrm{M}+\frac{g_\mathrm{M}^2}{\Gamma_\mathrm{M}(N)}\left[1-\cos\left(\frac{(2k+1)\pi}{N}\right)\right]\\
&\approx \Gamma_\mathrm{M}+\frac{g_\mathrm{M}^2}{\Gamma_\mathrm{M}},
\end{align}
\end{widetext}
where the first approximation uses $g_\mathrm{M}\ll \Omega$ and $\omega\approx\Omega_\mathrm{M}$, and the second one uses $\sum_{l=1}^{N-1}\left(1-\cos\left((2k+1)\pi/(N)\right)\right)\approx N$ with $N\gg 1$. By defining $g_\mathrm{M}=\eta\Gamma_\mathrm{M}$, we can derive the condition for the local parametric oscillation as
\begin{align}
\Gamma_\mathrm{M}(1+\eta^2)<\Gamma_\mathrm{OM} \label{eq4}
\end{align}

In a similar manner, we can determine the condition for the collective parametric oscillation, where a hybridized mechanical mode is collectively oscillating, by considering the hybridized mode in the whole mechanical node. This leads to the following coupled mode equations,
\begin{align}
\chi'_\mathrm{M}(\omega) h_l(\omega)=\sqrt{\frac{2}{N+1}}\sin\left(\frac{k\pi}{N+1}\right)F_\mathrm{OM},
\end{align}
where $h_l(\omega)$ is the hybridized displacement with $l=1$ to $N$, and $\chi'_\mathrm{M}(\omega)$ is the total susceptibility including the eigenvalue of the hybridized mode. Because the optomechanical gain is proportional to the square of the optomechanical coupling rate, it is renormalized to $\Gamma'_{\mathrm{OM},k}=\frac{2}{N+1}\sin^2\left(\frac{k\pi}{N+1}\right)\Gamma_\mathrm{OM}$. Thus, the condition for the global oscillation regime is given by
\begin{align}
\frac{N+1}{2}\Gamma_\mathrm{M}<\Gamma_\mathrm{OM}. \label{eq5}
\end{align} 

Which oscillations preferably occur in the device is determined by comparing Eqs. (\ref{eq4}) and (\ref{eq5}). Thus, an oscillation factor defined as
\begin{align}
\alpha_\mathrm{osc}=\frac{2(1+\eta_\mathrm{min}^2)}{N+1},
\end{align}
is used as an index for the parametric oscillation, where $\eta_\mathrm{min}$ is the minimum mechanical coupling factor among the multiple mechanical modes (RBMs). $\alpha_\mathrm{osc}<1$ reflects that the CMBR causes the local parametric oscillation; $\alpha_\mathrm{osc}>1$ reflects that it causes the collective parametric oscillation.

The coupling coefficient $\eta_{ij}$ between the $i$th-order axial mode in the microbottle A and $j$th-order axial mode in the microbottle B can be estimated by calculating the mode overlap integral, 
\begin{align}
\eta_{ij}=\frac{\sqrt{\Omega_\mathrm{A}\Omega_\mathrm{B}}}{\sqrt{\Gamma_\mathrm{A}\Gamma_\mathrm{B}}}\frac{\int\mathrm{d}V\vect{u}^{(i)}_\mathrm{A}(r,z)\cdot\vect{u}^{(j)}_\mathrm{B}(r,z)}{\sqrt{\int\mathrm{d}V |\vect{u}^{(i)}_\mathrm{A}(r,z)|^2}\sqrt{\int\mathrm{d}V |\vect{u}^{(j)}_\mathrm{B}(r,z)|^2}},
\end{align} 
where $\vect{u}^{(i)}_k(r,z)$, $\Omega_k$, and $\Gamma_k$ are the spatial distribution of the $i$th-order axial mode, mechanical resonance frequency, and damping factor of microbottle $k$, respectively ($k=$A, B). By utilizing the analytical expression of the mechanical modes \cite{asano2022free}, the minimum coupling coefficient between the 1st-order axial modes is determined to be $\eta_\mathrm{min}=\eta_{11}=0.5$. Thus, $\alpha_\mathrm{osc}=0.05<1$ reflects that the local parametric oscillation preferably occurs in the CMBR. 

For simplicity, the above discussion assumed that the optomechanical gain is independent of the mechanical modes. In practice, it depends on the mechanical frequencies through the laser detuning. For the 2nd-order axial mode, $\eta_{22}=7.8$, which is close to the experimental value of $\eta^{\mathrm{exp}}_{22}=6.9$ estimated from coupled mode theory \cite{asano2022free}. Because of $\alpha^\mathrm{2nd}_\mathrm{osc}=2.4>1$, a collective parametric oscillation might be induced by finely adjusting the laser detuning.

\section*{References}
\bibliographystyle{apsrev4-1}
\bibliography{manuscript}

\end{document}